\documentclass{article}
\usepackage{bm}
\begin{document}

\title{NEUTRINO MASS MATRIX SUBJECT TO $\mu-\tau$ SYMMETRY AND INVARIANT UNDER A CYCLIC PERMUTATION}

\maketitle

\begin{center}
\textbf Asan Damanik\footnote{d.asan@lycos.com}\\
Department of Physics\\Faculty of Science and Technology,\\Sanata Dharma University,\\Kampus III USD Paingan Maguwoharjo Sleman, Yogyakarta, Indonesia.
\end{center}
\begin{abstract}
Neutrino masses arise via a seesaw mechanism and its mass hierarchy, with assumption that heavy Majorana neutrino mass matrix subject to $\mu-\tau$ symmetry and invariant under a cyclic permutation, are evaluated.  Within this scenario, the neutrino masses: $\left|m_{1}\right|=\left|m_{2}\right|<\left|m_{3}\right|$ are obtained, which are incompatible with the experimental data.  By modifying neutrino mass matrix with the zero sum rule condition, the neutrino masses in inverted hierarchy: $\left|m_{3}\right|<\left|m_{1}\right|<\left|m_{2}\right|$ are obtained.

\begin{flushleft}
\end{flushleft}
\end{abstract}
\section{Introduction}

In the Standard Model of Electroweak interaction, it is impossible to understand the deficit of solar neutrino flux as measured in Earth.  For more than two decades solar neutrino flux measured on Earth has been much less than solar standard model prediction.\cite{Pantaleone91}  But, by using quantum mechanics concept, solar neutrino deficit can be explained if neutrino undergoes oscillation and mixing does exist in neutrino sector.  Neutrino oscillation implies that neutrinos have a non-zero mass.  Recently, there is a convincing evidence that neutrinos have a non-zero mass.  This evidence is based on the experimental facts that both solar and atmospheric neutrinos undergo oscillations.\cite{Fukuda98,Fukugita03}

A global analysis of neutrino oscillations data gives the best fit value to solar neutrino squared-mass differences\cite{Gonzales-Garcia04}:
\begin{equation}
\Delta m_{21}^{2}=(8.2_{-0.3}^{+0.3})\times 10^{-5}~{\rm eV^2}~
 \label{1}
\end{equation}
with
\begin{equation}
\tan^{2}\theta_{21}=0.39_{-0.04}^{+0.05},
 \label{2}
\end{equation}
and for the atmospheric neutrino squared-mass differences
\begin{equation}
\Delta m_{32}^{2}=(2.2_{-0.4}^{+0.6})\times 10^{-3}~{\rm eV^2}~
\end{equation}
with
\begin{equation}
\tan^{2}\theta_{32}=1.0_{-0.26}^{+0.35},
\end{equation}
where $\Delta m_{ij}^2=m_{i}^2-m_{j}^2~ (i,j=1,2,3)$ with $m_{i}$ are the neutrino mass eigenstates basis $\nu_{i}~(i=1,2,3)$ and $\theta_{ij}$ is the mixing angle between $\nu_{i}$ and $\nu_{j}$.  The mass eigenstates are related to the weak (flavors) eigenstates basis $(\nu_{e},\nu_{\mu},\nu_{\tau})$ as follows
\begin{equation}
\bordermatrix{& \cr
&\nu_{e}\cr
&\nu_{\mu}\cr
&\nu_{\tau}\cr}=V\bordermatrix{& \cr
&\nu_{1}\cr
&\nu_{2}\cr
&\nu_{3}\cr}
 \label{5}
\end{equation}
where $V$ is the mixing matrix.  The elements of the mixing matrix $V$ can be parameterized as\cite {Maki,Pontecorvo},
\begin{equation}
V=\bordermatrix{& & &\cr
&c_{12}c_{13} &s_{12}c_{13} &z^{*}\cr
&-s_{12}c_{23}-c_{12}z &c_{12}c_{23}-s_{12}s_{23}z &s_{23}c_{13}\cr
&s_{12}s_{23}-c_{12}c_{23}z &-c_{12}s_{23}-s_{12}c_{23}z &c_{23}c_{13}\cr},
 \label{aa}
\end{equation}
where $c_{ij}$ and $s_{ij}$ stand for $\cos\theta_{ij}$ and $\sin\theta_{ij}$ respectively, and $z=s_{13}e^{i\varphi}$.

In order to accommodate a non-zero neutrino mass-squared differences and the neutrino mixing, several models of neutrino mass together with neutrino mass generation have been proposed.\cite{Mohapatra98}  One of the most interesting mechanism which can generate a small neutrino mass is the seesaw mechanism, in which the right-handed neutrino $\nu_{R}$ has a large Majorana mass $M_{N}$ and the left-handed neutrino $\nu_{L}$ obtain a mass through leakage of the order of $~(m/M_{N})$ with $m$ the Dirac mass.\cite{Fukugita03}

According to seesaw mechanism\cite{Gell-Mann79}, the neutrino mass matrix $M_{\nu}$ is given by
\begin{equation}
M_{\nu}\approx -M_{D}M_{N}^{-1}M_{D}^T
 \label{Mnu}
\end{equation}
where $M_{D}$ and $M_{N}$ are the Dirac and Majorana mass matrices respectively.  The mass matrix model of a massive Majorana neutrino $M_{N}$ constrained by the solar and atmospheric neutrinos deficit and incorporate the seesaw mechanism and Peccei-Quinn symmetry have been reported by Fukuyama and Nishiura.\cite{Fukuyama97}  It has been a guiding principle that a presence of hierarchies or of tiny quantities imply a certain protection symmetry in undelying physics.  Candidates of such symmetry in neutrino physics may include $U(1)_{L'}$ based on the conservation of $L_{e}-L_{\mu}-L_{\tau}=L'$ and a $\mu-\tau$ symmetry based on the invariance of flavor neutrino mass term underlying the interchange of $\nu_{\mu}$ and $\nu_{\tau}$.  As we have already knew that the maximal mixing in atmospheric neutrino oscillation, as well as vanishing of the $U_{e3}$, are consecquencies of a $\mu-\tau$ symmetry.\cite{Lam,Fuki}  It is also pointed out by Ma\cite{Ma05} that it is more sense to consider the structure of $M_{N}$ for its imprint on $M_{\nu}$.

In order to consider the structure of the $M_{N}$ for its imprint on $M_{\nu}$, in this paper we construct a neutrino mass matrices via a seesaw mechanism.  We assumme that the heavy Majorana neutrino mass matrix $M_{N}$ subject to $\mu-\tau$ symmetry and invariant under a cyclyc permutation.  This paper is organized as follows: In Section 2, we determine the possible patterns for the heavy neutrino mass matrices $M_{N}$ subject to $\mu-\tau$ symmetry and invariant under a cyclic permutation.  The $M_{N}$ matrix which invariant under a cyclic permutation is used for obtaining the neutrino mass matrix $M_{\nu}$ via a seesaw mechanism.  In Section 3, we compare our neutrino mass matrix to the well-known neutrino mass matrices in literatures which known in agreement qualitatively with experimental data.  Finally, in Section 4 is devoted to a conclusion.

\section{The $\mu-\tau$ Symmetry and Invariant Under a Cyclic Permutation}

As we have explained in Section 1 that the aim of this paper is want to study the phenomenological consecquencies of neutrino mass matrix arise via a seesaw mechanism by imposing the requirement that the Majorana mass matrices subject to a $\mu-\tau$ symmetry and invariant under a cyclic permutation.  Thus, we first define the heavy Majorana neutrino mass matrix in Eq. (\ref{Mnu}) which can be parameterized by
\begin{equation}
M_{N}=\bordermatrix{& & &\cr
&A &B &C\cr
&B &D &E\cr
&C &E &F\cr}=\bordermatrix{& & &\cr
&M_{11} &M_{12} &M_{13}\cr
&M_{12} &M_{22} &M_{23}\cr
&M_{13} &M_{23} &M_{33}\cr}.
 \label{MN}
\end{equation}

Now, we are in position to impose the $\mu-\tau$ symmetry (also called $2-3$ symmetry) into heavy Majorana neutrino mass matrix in Eq. (\ref{MN}).  The $2-3$ symmetry is based on the invariance of the neutrino mass terms in the Lagrangian under the interchange of $2\leftrightarrow 3$ or $2\leftrightarrow -2$.  By imposing the $2-3$ symmetry into Eq. (\ref{MN}), we then have the heavy Majorana neutrino mass matrix in form
\begin{equation}
M_{N}=\bordermatrix{& & &\cr
&A &B &B\cr
&B &D &E\cr
&B &E &D\cr},
 \label{MN1}
\end{equation}
for the interchange of $2 \leftrightarrow 3$, and 
\begin{equation}
M_{N}=\bordermatrix{& & &\cr
&A &B &-B\cr
&B &D &E\cr
&-B &E &D\cr},
 \label{MN2}
\end{equation}
for the interchange of $2\leftrightarrow -3$.

In order to obtain the $M_{N}$ that invariant under a cyclic permutation among neutrino fields: $\nu_{1}\rightarrow \nu_{2}\rightarrow \nu_{3}\rightarrow \nu_{1}$, first we define
\begin{equation}
\nu'_{i}=U_{ij}\nu_{j},
 \label{T}
\end{equation}
where $U_{ij}$ are the matrix elements of a cyclic permutation matrix $U$.  Within the Eq. (\ref{T}), the $M_{N}$ satisfy
\begin{equation}
M'_{N}=U^{T}M_{N}U.
 \label{M}
\end{equation}
If $M_{N}$ matrix is invariant under a cyclic permutation, then the pattern of the $M'_{N}$ is the same with the $M_{N}$ pattern.

By checking the invariant forms of the $M_{N}$ matrices in Eqs. (\ref{MN1}) and (\ref{MN2}) together with the requirement that the $M_{N}$ are not singular matrices such that the $M_{N}$ has a $M_{N}^{-1}$, we found that there is no $M_{N}$ matrix invariant in form under a cycylic permutation.  But, by putting $A=D\neq 0$, $B=E\neq 0$ (case 1), $A=D\neq 0$, $B=E=0$ (case 2), and $A=D=0$, $B=E\neq 0$ (case 3) into Eq. (\ref{MN1}), then we can have the $M_{N}$ matrices to be invariant under a cyclic permutation. For the $M_{N}$ matrix in Eq. (\ref{MN2}), only the case 2 will give the neutrino mass matrix to be invariant under a cyclic permutation. Thus, we can have three patterns of heavy Majorana neutrino mass matrix $M_{N}$ subject to $\mu-\tau$ symmetry and invariant under a cyclic permutation with the patterns
\begin{equation}
M_{N}=\bordermatrix{& & &\cr
&A &B &B\cr
&B &A &B\cr
&B &B &A\cr},\ 
M_{N}=\bordermatrix{& & &\cr
&A &0 &0\cr
&0 &A &0\cr
&0 &0 &A\cr},\ 
M_{N}=\bordermatrix{& & &\cr
&0 &B &B\cr
&B &0 &B\cr
&B &B &0\cr}.
 \label{MN13}
\end{equation} 
From Eq.~(\ref{MN13}), we can see that the first matrix is the general mass matrix for heavy Majorana neutrino.  The second and the third mass matrices can be obtained from the first matrix by putting $B=0$ and $A=0$ respectively.  Thus, for further discussion, we only use the general heavy Majorana neutrino mass matrix pattern in Eq.~(\ref{MN13}).

The general heavy Majorana neutrino mass matrix can also be obtained from non-Abelian discrete group $\Delta(27)$ symmetry as noticed by Ma\footnote{Private discussion during the CTP International Conference on Neutrino Physics in the LHC Era, Luxor, Egypt, 15-10 November 2009.}.  The non-Abelian discrete group $\Delta(27)$ has 27 elements divided into 11 equivalence class.  It has 9 irreducible representation $\bf1_{i}$ ($i=1,...,9$) and 2 three dimensional ones $\bf3$ and $\bar{\bf3}$ with group multiplication\cite{Ma06}
\begin{eqnarray}
{\bf3}\otimes{\bar{\bf3}}=\sum^{9}_{i=1}{\bf1}_{i},\\
{\bf3}\otimes{\bf3}=\bar{\bf3}+\bar{\bf3}+\bar{\bf3},\\
\bar{\bf3}\otimes\bar{\bf3}={\bf3}+{\bf3}+{\bf3}.
\end{eqnarray}

Let the lepton doublets $(\nu_{i},l_{i})$ transform as $\bf3$ under $\Delta(27)$ and the lepton singlet $l^{c}_{i}$ as $\bar{\bf3}$, then with three Higgs doublets transforming as $\bf1_{1},\bf1_{2},\bf1_{3}$, the charged lepton and the Dirac neutrino mass matrix are diagonal and has three independent masses.  At the same time, with three Higgs triplets transforming as $\bf3$, the general form of heavy Majorana neutrino mass matrix in Eq.~(\ref{MN13}) is obtained when vacuum expectation values of three Higgs triplets are $(\bf1,\bf1,\bf1)$. 

\section{Neutrino Mass matrix via a Seesaw Mechanism}

To obtain a neutrino mass matrices $M_{\nu}$, we use the type-I seesaw mechanism in Eq. (\ref{Mnu}).  As we have shown in Section 2, the general heavy Majorana neutrino mass matrix $M_{N}$ and the Dirac neutrino mass matrix $M_{D}$ which are constructed by putting the neutrino mass matrices subject to $\mu-\tau$ symmetry and invariant under a cyclic permutation can also be obtained by using a non-Abelian discrete group $\Delta(27)$ symmetry.  The heavy Majorana and Dirac neutrino mass matrices to be used in the type-I seesaw mechanism are as following:
\begin{eqnarray}
M_{N}=\bordermatrix{& & &\cr
&A &B &B\cr
&B &A &B\cr
&B &B &A\cr},\label{mn}\\ 
M_{D}=\bordermatrix{& & &\cr
&m_{1}^{D} &0 &0\cr
&0 &m_{2}^{D} &0\cr
&0 &0 &m_{3}^{D}\cr}. \label{md}
\end{eqnarray}

By inserting neutrino mass matrices in Eqs.~(\ref{mn}) and (\ref{md}) into Eq.~(\ref{Mnu}), we then have
\begin{equation}
M_{\nu}=\frac{-1}{A^{2}+AB-2B^{2}}\bordermatrix{& & &\cr
&(A+B)(m_{1}^{D})^{2} &-Bm_{1}^{D}m_{2}^{D} &-Bm_{1}^{D}m_{3}^{D}\cr
&-Bm_{1}^{D}m_{2}^{D} &(A+B)(m_{2}^{D})^{2} &-Bm_{2}^{D}m_{3}^{D}\cr
&-Bm_{1}^{D}m_{3}^{D} &-Bm_{2}^{D}m_{3}^{D} &(A+B)(m_{3}^{D})^{2}\cr},
 \label{MN111}
\end{equation}
The pattern of neutrino mass matrix in Eq. (\ref{MN111}) has been proposed by Ma\cite{Ma03} and it compatible with the experimental results on neutrino oscillations.  Mohapatra and Rodejohann have also obtained the same pattern by using the strong scaling ansatz (SSA) concept.\cite{Mohapatra}  The same pattern of the $M_{\nu}$ in Eq. (\ref{MN111}) have also been obtained via a seesaw mechanism with heavy Majorana mass matrix subject to texture zero and invariant under a cyclic permutation as shown in Ref.\cite{Damanik}.  If we impose the $\mu-\tau$ symmetry into Eq. (\ref{MN111}), then we have $m_{2}^{D}=m_{3}^{D}$.  Putting $-(A+B)(m_{1}^{D})^{2}/(A^{2}+AB-2B^{2})=P$, $Bm_{1}^{D}m_{2}^{D}/(A^{2}+AB-2B^{2})=Q$, $-(A+B)(m_{2}^{D})^{2}/(A^{2}+AB-2B^{2})=R$, and $Bm_{2}^{D}m_{3}^{D}/(A^{2}+AB-2B^{2})=S$, we then have the neutrino mass matrix in form
\begin{eqnarray}
M_{\nu}=\bordermatrix{& & &\cr
&P &Q &Q\cr
&Q &R &S\cr
&Q &S &R\cr},
 \label{MN134}
\end{eqnarray}
which have been proposed in Ref.\cite{Ma05}.  The neutrino mass matrix of Eq. (\ref{MN134}) gives three eigenvalues, namely
\begin{eqnarray} 
m_{1,2}=\frac{1}{2}\left(P+R+S\mp \sqrt{P^{2}-2RP-2PS+R^{2}+2RS+S^{2}+8Q^{2}}\right),\nonumber\\ m_{3}=-S+R.
 \label{EV}
\end{eqnarray}

If the $M_{\nu}$ matrix in Eq. (\ref{MN134}) is diagonalized by $V$ in Eq. (\ref{5}) or Eq. (\ref{aa}) with $V$ is given by\cite{Ma05}
\begin{equation}
V=\bordermatrix{& & &\cr
&\cos\theta &-\sin\theta &0\cr
&\sin\theta/\sqrt{2} &\cos\theta/\sqrt{2} &-1/\sqrt{2}\cr
&\sin\theta/\sqrt{2} &\cos\theta/\sqrt{2} &1/\sqrt{2}\cr},
 \label{51}
\end{equation}
then we obtain,
\begin{equation}
\tan2\theta=\frac{4\sqrt{2}Q}{P-R-S}.
 \label{Te}
\end{equation}

If we use the experimental data in Eqs.~(\ref{1}) and (\ref{2}) and put $\theta_{12}=\theta$ , then the neutrino mass matrix gives the neutrino masses in hierarchy: $\left|m_{1}\right|= \left|m_{2}\right|<\left|m_{3}\right|$.  To accommodate the experimental data which show that $\left|m_{1}\right|\neq \left|m_{2}\right|$, therefore we use the zero sum rule condition: $m_{1}+m_{2}+m_{3}=0$ as an additional constraint on resulted neutrino mass matrix.  In this scheme, neutrino mass matrix still obey $\mu-\tau$ symmetry as well.

Using the zero sum rule condition, we then have $P=-2R$ and the neutrino mass matrix of Eq.~(\ref{MN134}) takes the form
\begin{equation}
M_{\nu}=\bordermatrix{& & &\cr
&-2R &Q &Q\cr
&Q &R &S\cr
&Q &S &R\cr}.
 \label{MN135}
\end{equation}
The neutrino mass matrix in Eq.~(\ref{MN135}) has the eigenvalues
\begin{eqnarray}
m_{1,2}=\frac{1}{2}\left[S-R\mp(3R+S)\sqrt{1+4\tan^{2}2\theta_{12}}\right],\nonumber\\ m_{3}=R-S.
 \label{EV1}
\end{eqnarray}
From Eq.~(\ref{EV1}), we can have the neutrino masses in inverted hierarchy: $\left|m_{3}\right|<\left|m_{1}\right|<\left|m_{2}\right|$ for $S>R$.

\section{Conclusion}
Neutrino mass matrix $M_{\nu}$ have been constructed via a seesaw mechanism with heavy Majorana neutrino mass matix subjects to a $\mu-\tau$ symmetry and invariant under a cyclic permutation.  In this scenario, the obtained neutrino mass matrix $M_{\nu}$ gives neutrino masses: $\left|m_{1}\right|= \left|m_{2}\right|<\left|m_{3}\right|$ which is incompatible with the experimental data.  Modifying the obtained neutrino mass matrix by zero sum rule condition, neutrino masses in inverted hierarchy: $\left|m_{3}\right|<\left|m_{1}\right|<\left|m_{2}\right|$ are obtained whose can predict squared-mass difference qualitatively.

\section*{Acknowledgments}

Author would like to thank to Prof. E. Ma and Prof S. Pakvasa for useful suggestions during the CTP INTERNATIONAL CONFERENCE ON NEUTRINO PHYSICS IN THE LHC ERA, 15-19 November 2009, Luxor, EGYPT.  This work is supported in part by Sanata Dharma University Yogyakarta and DP2M Ditjen Dikti Departemen Pendidikan Nasional Indonesia.

\end{document}